\documentclass[]{article}
\usepackage{arxiv}
\usepackage{graphicx} 
\usepackage{palatino} 
\usepackage{natbib}
\usepackage{booktabs}

\title{Using LLMs to Advance the Cognitive Science of Collectives}
\author{Ilia Sucholutsky\\
	New York University\\
	\texttt{is3060@nyu.edu} \\
	\And
	Katherine M. Collins \\
	University of Cambridge\\
	\texttt{kmc61@cam.ac.uk} \\
	\AND
	Nori Jacoby \\
	Cornell University \\
	\texttt{kj338@cornell.edu} \\
	\And
	Bill D. Thompson \\
	UC Berkeley \\
	\texttt{wdt@berkeley.edu} \\
	\And
	Robert D. Hawkins \\
	Stanford University \\
	\texttt{rdhawkins@stanford.edu } \\
}

\begin{document}

\maketitle

\textit{LLMs are already transforming the study of individual cognition, but their application to studying collective cognition has been underexplored. We lay out how LLMs may be able to address the complexity that has hindered the study of collectives and raise possible risks that warrant new methods. }

\section*{Introduction} 

Cognitive science and artificial intelligence (AI) have grown up together as fields.
The computational models of human minds developed in cognitive science have long served as benchmarks to articulate what it means for a system to be flexibly intelligent. Recent advances in AI, particularly around large language models (LLMs), are creating new opportunities to reciprocate this influence. 
Already, LLMs are being offered as scalable ``cognitive models'' of human behavior~\citep{binz2024centaur}, automatic analysts of unstructured psychological text~\citep{Rathje2024multilingual}, and components in neurosymbolic cognitive architectures (e.g., ~\citep{wong2023word}). 

However, most applications of LLMs to cognitive science have so far focused on \textit{individual} cognition. 
Here, we emphasize a comparatively underexplored frontier where LLMs are poised to transform cognitive science: the study of \textit{collective} cognition.
To see how powerful human interaction can be, imagine a single video call between a group of international research collaborators brainstorming a new project. Ahead of the conversation, the outcome seems near-impossible to predict, because the trajectory of ideas depends so heavily on each person’s current interests and expertise, their previous interactions, their cultural norms, what each person happens to think of, and how each person’s insights depend on the things the other people might say. 

\begin{figure*}[htb!]
    \centering
    \includegraphics[width=0.49\linewidth]{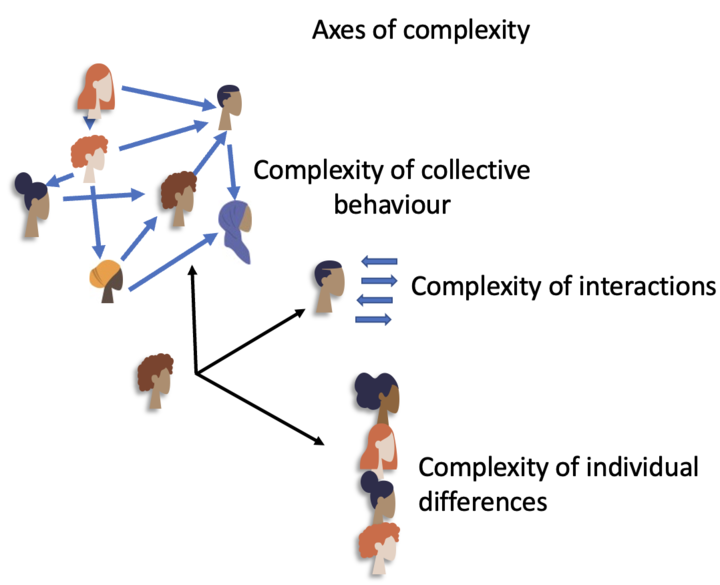}
    \includegraphics[width=0.49\linewidth]{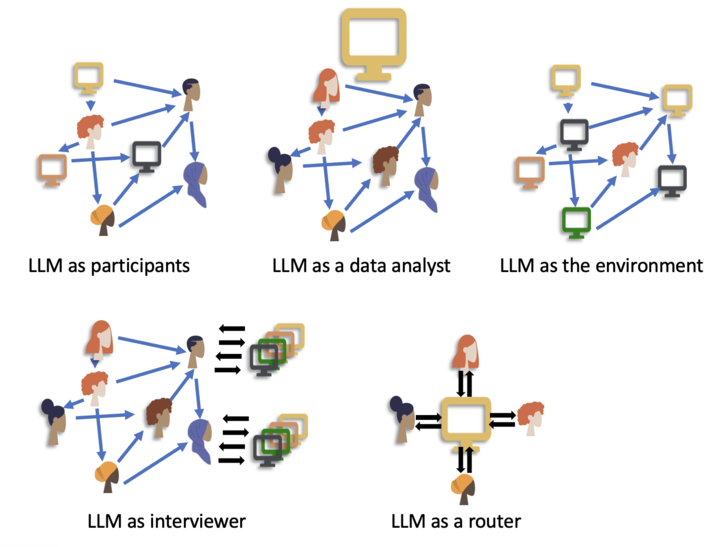}
    \caption{(\textbf{Left}) Axes of complexity in cognitive science. (i) Collective behavior, reflecting complexity at the network level (group structure and topology); (ii) Interactions, representing complexity at the edge level (connections within the social network); and (iii) Individual differences, capturing complexity at the node level (diversity among individuals). (\textbf{Right}) Roles of LLMs in cognitive science research. LLMs can serve at least five functions in studying collective behavior -— participant, data analyst, environment, interviewer, and router.
}
    \label{fig:axis-roles}
\end{figure*}

The emergent complexity of a single conversation is just the starting point for human social interactions. Our capacity for complex interaction is the reason we are able to cooperate in groups towards shared goals, to collaborate with other people across time and space, to communicate flexibly with strangers, and to accumulate new solutions to an open-ended range of conceptual and physical problems over generations. Social interaction in groups is fundamental to shaping human cognition and culture, but has been methodologically challenging for cognitive scientists to study at scale.
We propose a framework identifying three principal ``axes of complexity'' along which cognitive science research has struggled to scale up the study of complex networks of human and AI agents: (i) \emph{structural complexity}, reflecting larger and more varied networks of interacting agents (i.e. the topology of the social network), (ii) \emph{interactional complexity}, reflecting more sustained dyadic relationships (i.e. the edges of the network) that unfold over time, and (iii) \emph{individual complexity}, reflecting individual differences and cultural diversity in each agent's beliefs and behaviors (i.e. the nodes of the network). We summarize this Table~\ref{tab:complexity_axes} and Figure~\ref{fig:axis-roles} Left. 

We argue that LLMs can serve as powerful tools to address these axes of complexity, highlighting five possible roles LLMs can play in cognitive science: as \emph{participant}, \emph{interviewer}, \emph{environment}, \emph{router}, and \emph{data analyst} (see Table~\ref{tab:roles}; Figure~\ref{fig:axis-roles} Right). While some of these roles are already being employed to study individual cognition, we focus here on why and how LLMs are particularly well-positioned to advance the cognitive science of collectives along all three axes of complexity -- and serve as a particularly well-suited domain where computational scientists and cognitive scientists are well-positioned to collaborate. 
We concretely illustrate this potential by highlighting a few examples of recent research utilizing LLMs to study collective intelligence before discussing emerging risks and capability gaps, and future research directions to mitigate these.

\begin{table}[tb!]
\centering
\begin{tabular}{p{6em}|p{25em}|p{10.5em}}
\toprule
\textbf{Axis} & \textbf{Description} & \textbf{Example} \\
\midrule
Structural Complexity & Recent cognitive science research highlights how increasingly complex social networks involving humans and large language models (LLMs) produce emergent behaviors distinct from dyadic interactions. & \cite{brinkmann2023machine} \\

Interactional Complexity & Interactions between humans and LLMs are becoming richer, encompassing diverse modalities (audio, video, gestures), greater reciprocity (e.g., human-in-the-loop computation), and increasingly sophisticated representations. Machines act as "thought partners," continuously inferring human states and goals, mirroring human-human dynamics. & 
\cite{collins2024building} \\

Individual Complexity & Cognitive science has long recognized individual differences within and across societies. These differences enhance interaction diversity with personalized AI systems. LLMs evaluated in multiple languages can capture cultural variations, and personalized LLMs simulate distinct individual behaviors. & \cite{marjieh2024large}\hfill  \, \, \cite{park2024generative} \\
\bottomrule
\end{tabular}
\caption{Three axes of cognitive complexity, their descriptions, and examples from the literature that already leverage LLMs or other AI systems.}
\label{tab:complexity_axes}
\end{table}

\begin{table}[htb!]
    \centering
    \begin{tabular}{p{6em}|p{25em}|p{10.5em}}
    \toprule
      \textbf{Role}   &  \textbf{Description} & \textbf{Example}\\
      \midrule
       Participant & LLM acts as a participant in a task, directly interacting with human participants and other AI agents embedded in the social network. &\cite{marjieh2024large} \\
        Interviewer & LLM acts as an interviewer, asking questions and eliciting interactive textual data from human participants.  &\cite{park2024generative}\\
        Environment &LLM simulates an interactive environment where multiple players engage with a coherently generated world (e.g., as a Game Master in a text-based adventure).& \cite{gallotta2024large}\\
        Router & LLM summarizes inputs from multiple human participants and routes information effectively to facilitate emergent consensus.& \cite{tessler2024habermas} \\
        Analyst &LLM annotates and analyzes data from large-scale collective behavior datasets (e.g., social network data.). Crucially, LLMs can automatically process large unstructured datasets into quantified, analyzable forms without the need for human annotations.  & \cite{Rathje2024multilingual}\\
        \bottomrule
    \end{tabular}
    \caption{Five roles of LLMs for studies of collective cognition, their descriptions, and examples from the literature.}
    \label{tab:roles}
\end{table}



\section*{Unrealized potential of LLMs for studying collective behavior
} 

Much of the richness and open-endedness of human intelligence arises not from individual brain-power alone, but from our interactions with each other. Our distinctive capacities for interaction, collaboration, and knowledge accumulation are deeply cognitive problems. However, multi-agent problems such as these have traditionally been difficult to study. Experimental studies of multi-agent interaction have historically been difficult to orchestrate and costly to conduct. Similarly, the development of formal cognitive models of interacting systems of agents has historically been challenging to pursue, and typically limited to highly simplified construals of human learning and reasoning. 

New methods in cognitive science are breaking through these barriers. For example, new technologies for large-scale online participant recruitment, combined with emerging approaches to algorithmic experimental design, have enabled substantial progress in the experimental science of interactive intelligence. Similarly, more powerful and flexible methods for developing and training richly structured cognitive models are rapidly expanding our capacity to construct cognitively-meaningful formal theories of interacting minds. Nonetheless, these advances remain limited in fundamental ways, such as cost and scaling limits on participant recruitment, and the brittleness of domain-specificity in traditional cognitive models.
LLMs offer important opportunities to potentially overcome several of these key limitations from potentially acting as simulated participants to helping acting as a dynamic environment for human-human participants. 

\section*{Taking on the axes of complexity with LLMs} 

We next spotlight a few existing works that begin to demonstrate the potential power of using LLMs to study collective behavior. We highlight the role that LLMs play in each and how the methodology can enable other studies of collective cognition. 

Consider \cite{tessler2024habermas}, which employed an iterative approach where human statements on contested issues (e.g., abortion, immigration) were summarized by an LLM (i.e., LLM as router), then ranked and critiqued by humans, resulting in consensus. This method increased interaction complexity and potentially individual diversity (pool of over 5,000 individuals), and showcases the potential value of using LLMs to enable humans to engage with each other in new streamlined ways, at scale. While the core focus in this work is on the use of LLMs to foster new kinds of human-LLM societal interactions (e.g., to improve deliberation), the methods pioneered here involving the use of LLMs to summarize and communicate across many people at once could be used to advance the study of collectives --- e.g., humans in diverse collectives reach consensus on decisions that affect them all. However, we note that, at present, the study was limited along the structural complexity axis, featuring only a linear chain of six alternating LLM-human stages. 


In contrast, a recent study by \cite{shiiku2025dynamics} enhanced collective complexity by placing 625 humans or LLM agents within an experimental social network, wherein participants selected and creatively rewrote stories received from their neighbors in the network. Although interactions with LLMs were relatively simple (LLMs acted as participants performing story selection), the experiment highlighted the complexity of the network topology and demonstrated greater diversity within hybrid human-AI groups compared to purely human or purely AI societies. This extends recent cognitive science research that embeds participants within experimental social networks, exploring complex interactions such as algorithmic sorting \citep{thompson2022complex}. 

\begin{figure}[htb!]
    \centering
    \includegraphics[width=0.5\linewidth]{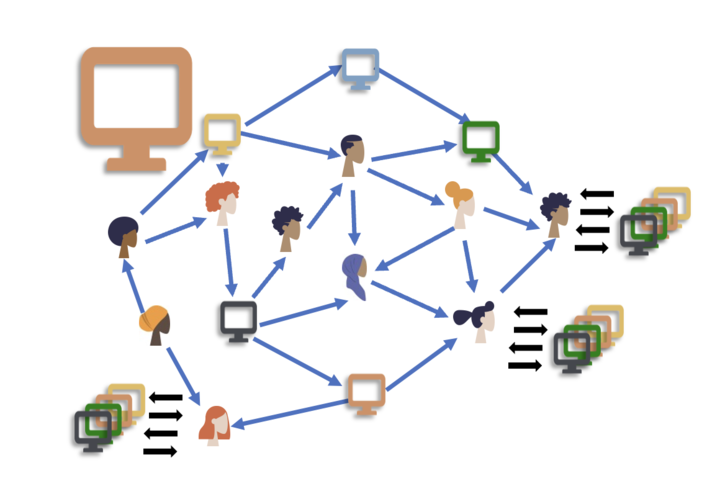}
    \caption{Future experiments can explore integrating the three axes of complexity to raise and address previously out-of-reach questions in collective cognition.}
    \label{fig:intial-and-future-direction}
\end{figure}

Both of these studies leveraged LLMs to scale along \textit{one} of the three axes of complexity that we lay out in our framework, demonstrating the initial promise of utilizing LLMs for grappling with such complexity. Future research can explore ways to use LLMs to jointly address multiple axes at once (see Fig. \ref{fig:intial-and-future-direction}), for example, in studying how cultural knowledge, like algorithms, flow and evolve across many generations in large societies \citep{thompson2022complex}, or how people adapt their communicative norms at scales ranging from dyadic communication (e.g., debating the merits of a policy with another person) all the way up to society-level communication (e.g., delivering a persuasive presidential campaign speech) \citep{Hawkins2023-HAWFPT-2}. Conducting such large-scale studies is especially pressing today to help us understand how we engage with each other across increasingly wide scales (a single interaction, a local constituency, a country, and across countries) and enable us to forecast how changes to this social fabric may be -- and even already are --  being transformed by the rapid proliferation of AI systems in our daily lives~\citep{brinkmann2023machine}. 




\section*{Missing capabilities and usage risks}
While we lay out an optimistic vision of how LLMs can advance the study of collective cognition, we note that achieving this vision is not simply a matter of convincing cognitive scientists to use new models in their experiments. There still remain a number of capability gaps and usage risks that need to be addressed to catalyze widespread adoption and deep integration of LLMs and enable the kind of studies we envision above. We lay out several open challenges below that we believe can both be solved in the near term and have a high impact on advancing a cognitive science of collectives. It is worth noting that the challenges we outline are not exclusive to cognitive science, and we hope that this encourages LLM researchers and computational scientists to take them on.

\paragraph{Interpretability, transparency, and alignment}
To leverage LLMs as participants, models of cognition, or thought partners, we need these systems to be well-aligned with human representations and behavior~\citep{marjieh2024large, binz2024centaur}. Finding the best ways to measure and increase this type of alignment is a significant open challenge~\citep{sucholutsky2023getting}. 
Increased interpretability and transparency at both the representational and algorithmic levels would also be highly beneficial.
For example, if considering using LLMs as \textit{components} of cognitive models or networks of collective intelligence, it is not immediately clear what their ``cognitive meaning'' is if we do not know what the inner workings of the model are, nor what data they may have been trained on. 

\paragraph{Cultural representation gaps}

LLMs can harbor significant biases, predominantly reflect the communicative norms, group dynamics, and social structures of the dominant cultural contexts in which they were trained \citep{ryan2024unintended}, and can only roughly approximate the linguistic and cross-cultural diversity of groups of participants from around the world~\citep{marjieh2024large}. 
When studying collective intelligence across cultural contexts, these representation gaps become particularly problematic, as they may mischaracterize how different communities cooperate, deliberate, and build shared knowledge. These biases threaten the ecological validity of group cognition research. 

\paragraph{Homogenization} 


LLMs tend to homogenize the diversity of human behavior and interaction patterns. Despite attempts to introduce variability through ``personas'' or character prompting \citep{park2024generative}, LLMs and other ``human-like'' AI systems often produce a narrower distribution of responses than would be observed in actual human populations~\citep{meister2024benchmarking}. 
This homogenization is particularly problematic for collective studies, where emergent group phenomena depend critically on the heterogeneity of individual actors, their diverse knowledge bases, opinions, and communication patterns. 

\paragraph{Reproducibility} Another risk to incorporating LLMs as components in cognitive science research is their potential lack of reproducibility. There is a risk to researchers tailoring their research to any one model in particular. The model (if hosted via an API) could become deprecated (as has happened with the Codex variety), rendering a particular strain of research no longer reproducible. Or, as the pace of development around LLMs continues to evolve at breakneck pace, a model that researchers may consider ``state of the art'' may rapidly change. To that end, we encourage researchers who are considering using LLMs as a component or tool in the study of collective intelligence to ensure their study design is \textit{robust} to the particular choice of LLM and ideally modular to permit swapping in and out LLMs, and that the choice of which LLMs to consider takes into account the necessity of interpretability depending on the role the LLM aims to fill. 

\paragraph{Compute costs}

Finally, scaling LLM research to study collective cognition introduces exponentially greater computational demands compared to individual cognitive modeling. While simulating a single agent requires one model instance, modeling multi-agent interactions potentially requires $O(n^2)$ model calls to represent all pairwise interactions in a group of $n$ agents. This scaling becomes particularly problematic for larger collectives or longitudinal studies. The resulting computational costs are not merely economic but represent significant environmental impacts through increased energy consumption and carbon emissions. 
Additionally, the computational intensity of these approaches risks creating research inequities, where only well-resourced institutions can conduct group cognition studies at scale. 

\section*{Looking ahead}
In just a few years, LLMs have already made important impacts on methodology and theory in cognitive science, shedding new light on longstanding questions about the role of language in human cognition and cognitive development, and raising exciting new questions around what can be learned from language alone. In the years to come, we hope that the advancement of LLMs will support similar innovations in our understanding of multi-agent intelligence. Computational scientists have a role to play in shaping LLMs for the study of collective cognition, too, to address the risks and capability gaps that we laid out above. Of course, we note that LLMs are just one tool in a burgeoning toolkit of computational methods that can open up new frontiers in cognitive science, but they need not be the only tool.  

We close by calling for more joint work by both computational scientists and cognitive scientists in scaling research along all three axes of complexity. 
To scale \textit{structural complexity},
researchers could explore using well-established experiment tools from computational cognitive science, such as dallinger\footnote{www.dallinger.dev} and PsyNet\footnote{www.psynet.dev}, to jointly orchestrate large groups of both LLMs and human participants. As scientific tools, LLMs have much to contribute to the development of the infrastructure we rely on to conduct interactive and large-scale experimental studies. 
To scale \textit{interactional complexity}, LLMs need to become effective interviewers and not just rote analyzers but analyzers capable of collaborative sensemaking with cognitive science researchers. We encourage more work to build genuine AI thought partners~\citep{collins2024building} grounded in computational cognitive science. 
Finally, to capture \textit{individual complexity}, we encourage researchers to train LLMs that reflect human variability across individuals and cultures. Moreover, benchmarks should extend beyond evaluating accuracy in coding and math problems to assessing LLMs’ ability to capture the diverse spectrum of human opinions and preferences~\citep{ying2025benchmarking}. The axes of complexity we lay out here are thorny -- requiring rich and creative innovation that draws on a range of computational methods, old and new, and is a domain ripe for further collaboration between computational scientists and cognitive scientists to truly understand the richness that make up our interactions with each other. 







\bibliographystyle{plainnat}
\bibliography{main}

\begin{thebibliography}{16}
\providecommand{\natexlab}[1]{#1}
\providecommand{\url}[1]{\texttt{#1}}
\expandafter\ifx\csname urlstyle\endcsname\relax
  \providecommand{\doi}[1]{doi: #1}\else
  \providecommand{\doi}{doi: \begingroup \urlstyle{rm}\Url}\fi

\bibitem[Binz et~al.(2024)Binz, Akata, Bethge, Br{\"a}ndle, Callaway, Coda-Forno, Dayan, Demircan, Eckstein, {\'E}ltet{\H{o}}, et~al.]{binz2024centaur}
Marcel Binz, Elif Akata, Matthias Bethge, Franziska Br{\"a}ndle, Fred Callaway, Julian Coda-Forno, Peter Dayan, Can Demircan, Maria~K Eckstein, No{\'e}mi {\'E}ltet{\H{o}}, et~al.
\newblock Centaur: a foundation model of human cognition.
\newblock \emph{arXiv preprint arXiv:2410.20268}, 2024.

\bibitem[Brinkmann et~al.(2023)Brinkmann, Baumann, Bonnefon, Derex, M{\"u}ller, Nussberger, Czaplicka, Acerbi, Griffiths, Henrich, et~al.]{brinkmann2023machine}
Levin Brinkmann, Fabian Baumann, Jean-Fran{\c{c}}ois Bonnefon, Maxime Derex, Thomas~F M{\"u}ller, Anne-Marie Nussberger, Agnieszka Czaplicka, Alberto Acerbi, Thomas~L Griffiths, Joseph Henrich, et~al.
\newblock Machine culture.
\newblock \emph{Nature Human Behaviour}, 7\penalty0 (11):\penalty0 1855--1868, 2023.

\bibitem[Collins et~al.(2024)Collins, Sucholutsky, Bhatt, Chandra, Wong, Lee, Zhang, Zhi-Xuan, Ho, Mansinghka, et~al.]{collins2024building}
Katherine~M Collins, Ilia Sucholutsky, Umang Bhatt, Kartik Chandra, Lionel Wong, Mina Lee, Cedegao~E Zhang, Tan Zhi-Xuan, Mark Ho, Vikash Mansinghka, et~al.
\newblock Building machines that learn and think with people.
\newblock \emph{Nature Human Behaviour}, 8\penalty0 (10):\penalty0 1851--1863, 2024.

\bibitem[Gallotta et~al.(2024)Gallotta, Todd, Zammit, Earle, Liapis, Togelius, and Yannakakis]{gallotta2024large}
Roberto Gallotta, Graham Todd, Marvin Zammit, Sam Earle, Antonios Liapis, Julian Togelius, and Georgios~N Yannakakis.
\newblock Large language models and games: A survey and roadmap.
\newblock \emph{IEEE Transactions on Games}, 2024.

\bibitem[Hawkins et~al.(2023)Hawkins, Franke, Frank, Goldberg, Smith, Griffiths, and Goodman]{Hawkins2023-HAWFPT-2}
Robert~D. Hawkins, Michael Franke, Michael~C. Frank, Adele~E. Goldberg, Kenny Smith, Thomas~L. Griffiths, and Noah~D. Goodman.
\newblock From partners to populations: A hierarchical bayesian account of coordination and convention.
\newblock \emph{Psychological Review}, 130\penalty0 (4):\penalty0 977--1016, 2023.
\newblock \doi{10.1037/rev0000348}.

\bibitem[Marjieh et~al.(2024)Marjieh, Sucholutsky, van Rijn, Jacoby, and Griffiths]{marjieh2024large}
Raja Marjieh, Ilia Sucholutsky, Pol van Rijn, Nori Jacoby, and Thomas~L Griffiths.
\newblock Large language models predict human sensory judgments across six modalities.
\newblock \emph{Scientific Reports}, 14\penalty0 (1):\penalty0 21445, 2024.

\bibitem[Meister et~al.(2024)Meister, Guestrin, and Hashimoto]{meister2024benchmarking}
Nicole Meister, Carlos Guestrin, and Tatsunori Hashimoto.
\newblock Benchmarking distributional alignment of large language models.
\newblock \emph{arXiv preprint arXiv:2411.05403}, 2024.

\bibitem[Park et~al.(2024)Park, Zou, Shaw, Hill, Cai, Morris, Willer, Liang, and Bernstein]{park2024generative}
Joon~Sung Park, Carolyn~Q Zou, Aaron Shaw, Benjamin~Mako Hill, Carrie Cai, Meredith~Ringel Morris, Robb Willer, Percy Liang, and Michael~S Bernstein.
\newblock Generative agent simulations of 1,000 people.
\newblock \emph{arXiv preprint arXiv:2411.10109}, 2024.

\bibitem[Rathje et~al.(2024)Rathje, Mirea, Sucholutsky, Marjieh, Robertson, and Bavel]{Rathje2024multilingual}
Steve Rathje, Dan-Mircea Mirea, Ilia Sucholutsky, Raja Marjieh, Claire~E. Robertson, and Jay J.~Van Bavel.
\newblock {GPT} is an effective tool for multilingual psychological text analysis.
\newblock \emph{Proceedings of the National Academy of Sciences}, 121\penalty0 (34):\penalty0 e2308950121, 2024.

\bibitem[Ryan et~al.(2024)Ryan, Held, and Yang]{ryan2024unintended}
Michael Ryan, William Held, and Diyi Yang.
\newblock Unintended impacts of {LLM} alignment on global representation.
\newblock In \emph{Proceedings of the 62nd Annual Meeting of the Association for Computational Linguistics (Volume 1: Long Papers)}, pages 16121--16140, 2024.

\bibitem[Shiiku et~al.(2025)Shiiku, Marjieh, Anglada-Tort, and Jacoby]{shiiku2025dynamics}
Shota Shiiku, Raja Marjieh, Manuel Anglada-Tort, and Nori Jacoby.
\newblock The dynamics of collective creativity in human-ai social networks.
\newblock \emph{arXiv preprint arXiv:2502.17962}, 2025.

\bibitem[Sucholutsky et~al.(2023)Sucholutsky, Muttenthaler, Weller, Peng, Bobu, Kim, Love, Grant, Groen, Achterberg, et~al.]{sucholutsky2023getting}
Ilia Sucholutsky, Lukas Muttenthaler, Adrian Weller, Andi Peng, Andreea Bobu, Been Kim, Bradley~C Love, Erin Grant, Iris Groen, Jascha Achterberg, et~al.
\newblock Getting aligned on representational alignment.
\newblock \emph{arXiv preprint arXiv:2310.13018}, 2023.

\bibitem[Tessler et~al.(2024)Tessler, Bakker, Jarrett, Sheahan, Chadwick, Koster, Evans, Campbell-Gillingham, Collins, Parkes, Botvinick, and Summerfield]{tessler2024habermas}
Michael~Henry Tessler, Michiel~A. Bakker, Daniel Jarrett, Hannah Sheahan, Martin~J. Chadwick, Raphael Koster, Georgina Evans, Lucy Campbell-Gillingham, Tantum Collins, David~C. Parkes, Matthew Botvinick, and Christopher Summerfield.
\newblock {AI} can help humans find common ground in democratic deliberation.
\newblock \emph{Science}, 386\penalty0 (6719):\penalty0 eadq2852, 2024.

\bibitem[Thompson et~al.(2022)Thompson, Van~Opheusden, Sumers, and Griffiths]{thompson2022complex}
Bill Thompson, B~Van~Opheusden, T~Sumers, and TL~Griffiths.
\newblock Complex cognitive algorithms preserved by selective social learning in experimental populations.
\newblock \emph{Science}, 376\penalty0 (6588):\penalty0 95--98, 2022.

\bibitem[Wong et~al.(2023)Wong, Grand, Lew, Goodman, Mansinghka, Andreas, and Tenenbaum]{wong2023word}
Lionel Wong, Gabriel Grand, Alexander~K Lew, Noah~D Goodman, Vikash~K Mansinghka, Jacob Andreas, and Joshua~B Tenenbaum.
\newblock From word models to world models: Translating from natural language to the probabilistic language of thought.
\newblock \emph{arXiv preprint arXiv:2306.12672}, 2023.

\bibitem[Ying et~al.(2025)Ying, Collins, Wong, Sucholutsky, Liu, Weller, Shu, Griffiths, and Tenenbaum]{ying2025benchmarking}
Lance Ying, Katherine~M Collins, Lionel Wong, Ilia Sucholutsky, Ryan Liu, Adrian Weller, Tianmin Shu, Thomas~L Griffiths, and Joshua~B Tenenbaum.
\newblock On benchmarking human-like intelligence in machines.
\newblock \emph{arXiv preprint arXiv:2502.20502}, 2025.

\end{thebibliography}

\end{document}